\documentclass[prd,amsmath,amssymb]{revtex4}

\usepackage{graphicx}
\usepackage{dcolumn}
\usepackage{bm}

\def\Journal#1#2#3#4{{#1} {\bf #2}, #3 (#4)}

\def\NIMA{{\em Nucl. Instrum. Methods} A}

\def\PRL{\em Phys. Rev. Lett.}
\def\PRD{{\em Phys. Rev.} D}

\def\JPG{{\em J. Phys.} G}

\def\be{\begin{equation}}
\def\ee{\end{equation}}
\def\bea{\begin{eqnarray}}
\def\eea{\end{eqnarray}}

\begin{document}
\noindent\begin{minipage}{0.3\textwidth}
\end{minipage}%
\hfill%
\begin{minipage}{0.6\textwidth}\raggedleft
FERMILAB-CONF-14-336-E\\
Proceedings for Moriond QCD 2014 \\
May 15, 2014
\end{minipage}

\vspace*{1.5cm}

\title{MEASUREMENTS OF TOP QUARK PRODUCTION AND PROPERTIES AT THE TEVATRON}

\author{P. BARTO\v{S} on behalf of CDF and D\O\ Collaborations}

\affiliation{\vspace*{0.2cm} Department of nuclear physics and biophysics, Faculty of Mathematics, Physics and Informatics, \\
Comenius University, Mlynska Dolina F1, 842 48 Bratislava, Slovakia}

\begin{abstract}
\vspace*{2.0cm}

In this letter, we summarize the latest results of the top-quark production and properties at the Tevatron. We do not include results of the top-quark mass and single top-quark production as they were presented in separate talks. The results of the measurements are mostly consistent with the standard-model predictions. However, by looking at the production asymmetry measured by CDF, one can see a discrepancy in both, $t\bar{t}$ inclusive and lepton-based measurements. D\O\ results of production asymmetry are compatible with the standard-model predictions as well as with the CDF results.

\end{abstract}


\maketitle

\section{Introduction}
The top quark is the heaviest fundamental particle with the unique properties. It is the 3$^{rd}$ generation up-type quark with electric charge of +2/3$e$ and mass of $173.34 \pm 0.76$\ GeV/$c^{2}$ ~\cite{WorldComb}. The huge mass means that the top quark is an excellent perturbative object for testing QCD as it is produced at small distances ($\sim 1/m_\mathrm{top}$) characterized by low value of coupling constant $\alpha_\mathrm{S}\approx 0.1$.
 Due to the very short lifetime ($\sim 10^{-25}$\ s), the top quark decays before hadronization and we can study its properties using its decay products. If a deviation of the measured properties from the Standard model (SM) predictions is seen, it could be a sign of a new physics. The top-quark events are important background in the Higgs-boson studies.

At the Tevatron $p\bar{p}$ collider, top quarks are produced mainly in pairs through strong force quark-antiquark annihilation ($\sim 85$\%) and gluon-gluon fusion ($\sim 15$\%) processes. According the SM, the top quark decays into the $W$ boson and bottom ($b$) quark in almost 100\% of the cases. The final state of top-quark-pair production contains two $b$-quarks jets and two $W$ bosons, which decay leptonically (to $l\nu_{l}$, where in our case $l = e,\mu$) or hadronically (into quarks). The $t\bar{t}$ events can be then classified into three categories: the $dilepton$ or $all$--$jets$ events, where both $W$ bosons decay leptonically or hadronically, respectively; and the $lepton$$+$$jets$ events, where one of the $W$ bosons decays leptonically while the other one decays hadronically.

\section{Top-quark pair production cross section}
\subsection{Inclusive cross section}\label{subsec:inclxsec}

Using the BLUE method~\cite{blue}, we combine two D\O\ and four CDF measurements with weights of $40\%$ and $60\%$, respectively. The obtained value of the cross section is $\sigma_{t\bar{t}} = 7.60 \pm 0.41$ pb~\cite{xsec_comb} for a top-quark mass of $172.5$ GeV/$c^2$. The contributions to the uncertainty are 0.20 pb from statistical sources, 0.29 pb from systematic sources, and 0.21 pb from the uncertainty on the integrated luminosity. The main source of the systematic uncertainty is signal modeling. The result is in good agreement with the SM prediction of $7.35^{+0.28}_{-0.33}$ pb at NNLO+NNLL in pertubative QCD~\cite{SMxsec}.
 
\subsection{Differential cross section}\label{subsec:diffxsec}

Measurement of $t\bar{t}$ differential cross section provides direct test of QCD. Measurements deepen our understanding of QCD, and help us to improve the modeling of QCD processes. A precise QCD modeling is vital for many new-physics searches, where the differential top-quark cross sections are used to constrain new sources of physics (e.g. axigluon models).

D\O\ presents a measurement using full data sample ($9.7$ fb$^{-1}$) of $lepton$$+$$jets$ events. The differential cross section is studied as a function of the transverse momentum and absolute value of the rapidity of the top quarks as well as of the invariant mass of the $t\bar{t}$ pair. The events are required to contain an isolated lepton, a large imbalance in transverse momentum, and four or more jets with at least one jet identified to originate from a $b$ quark. The final state is obtained by kinematic reconstruction and the result is corrected to parton-level top quark. 
The observed differential cross sections~\cite{D0diffxsec} are consistent with SM predictions. 
The authors compared the measurements also with predictions, which include different axigluon models with different couplings provided by Falkowski {\it et al}.~\cite{Falkowski}. One can say that some models are in tension with the presented data. 

CDF reports a measurement of the differential cross section as a function of the top-quark production angle. The measurement is performed using $lepton$$+$$jets$ events, corresponding to an integrated luminosity of $9.4$ fb$^{-1}$. The authors employ the Legendre polynomials to characterize the shape of the differential cross section at the parton level. 
As Legendre moment $a_{0}$ contains only total cross section, all Legendre moments of degree $l>0$ ($a_l$) are scaled, so that $a_{0}=1$. The observed Legendre moments are in good agreement with the NLO SM predictions~\cite{LegendreMomPredict}, with the exception of $\sim 2\sigma$ excess linear-term moment $a_{1}=0.40 \pm 0.12$ ~\cite{CDFdiffxsec}, compared to standard-model prediction of $a_{1}=0.40^{+0.07}_{-0.03}$. The result constrains t-channel explanations of the $t\bar{t}$ forward-backward asymmetry (see section~\ref{sec:asym}) and favors asymmetry models with strong s-channel components (i.e. axigluon models).

\section{Decay width and lifetime}\label{sec:width}
As we already mentioned, top quark is the heaviest elementary particle and its large mass endows it with the largest decay width. A deviation from the SM predictions could indicate decays via e.g., charged Higgs boson, stop squark, or flavor changing neutral current. 

CDF reports direct measurement using $t\bar{t}$-pair candidates reconstructed in the final state with one charged lepton and at least four hadronic jets, corresponding to full Tevatron Run II data set ($8.7$ fb$^{-1}$). The total decay width is extracted by comparing reconstructed top-quark mass and mass of the hadronically decaying $W$ boson with distributions derived from simulated signal and background samples. For the top-quark mass of $172.5$ GeV/$c^2$, CDF obtains $\Gamma_{t}=2.21^{+1.84}_{-1.11}$ GeV, what corresponds to a lifetime of $1.6 \times 10^{-25} < \tau_{t} < 6.0 \times 10^{-25}$ s ~\cite{CDFwidth}. The results is in agreement with the SM expectations.

The D\O\ most precise measurement is based on extracting of total decay width from the partial decay width $\Gamma(t\rightarrow Wb)$ and the branching fraction ${\cal B} (t\rightarrow Wb)$. The $\Gamma(t\rightarrow Wb)$ is obtained from the $t$-channel single top-quark production cross section and ${\cal B} (t\rightarrow Wb)$ is measured in $t\bar{t}$ events. The method assumes that the coupling leading to $t$-channel single top-quark production is identical to the coupling in the top-quark decay. For the top-quark mass of $172.5$ GeV/$c^2$, the resulting width is $\Gamma_{t}=2.00^{+0.47}_{-0.43}$ GeV, what corresponds to a lifetime of $\tau_{t} = 3.29^{+0.90}_{-0.63} \times 10^{-25}$ s  ~\cite{D0width}.

\section{Branching fractions}\label{sec:br}
The decay rate of top quark into a $W$ boson and a down-type quark $q$ ($q=d,s,b$) is proportional to $|V_{tq}|^2$, the squared module of element of CKM matrix. Under the assumption of a unitary $3 \times 3$ CMK matrix, $|V_{tb}|$ is highly constrained to $|V_{tb}| = 0.999152^{+0.000030}_{-0.000045}$~\cite{Vtb_pred}, and the top quark decays almost exclusively to $Wb$. The existence of a fourth generation of quarks would remove this constraint, would lead to smaller values of $|V_{tb}|$, and could also affect the decay rates in the $t\bar{t}$ production. The latter can be used to extract the ratio of branching fractions, $R= {\cal B}(t \rightarrow Wb)/{\cal B}(t \rightarrow Wq)$. 

CDF reports the measurement in the $dilepton$ decay channel using 8.7 fb$^{-1}$ of data. The ratio obtained by the maximum likelihood estimator has value of $R = 0.87 \pm 0.07$. Assuming the unitarity of the CKM matrix and the existence of the three quark generations, the authors extract $|V_{tb}| = 0.93 \pm 0.04$~\cite{CDF_Vtb}.

The D\O\ measurement uses data sample of 5.4 fb$^{-1}$. The final result is obtained as combination of $lepton+jets$- and $dilepton$-channel results. The obtained value of the ratio is $R=0.90 \pm 0.04$, and the extracted CKM matrix element $|V_{tb}| = 0.95 \pm 0.02$, assuming unitarity of the $3 \times 3$ CMK matrix~\cite{D0_Vtb}.

\begin{figure}
\begin{minipage}{0.48\linewidth}
\centering{\includegraphics[width=0.95\linewidth]{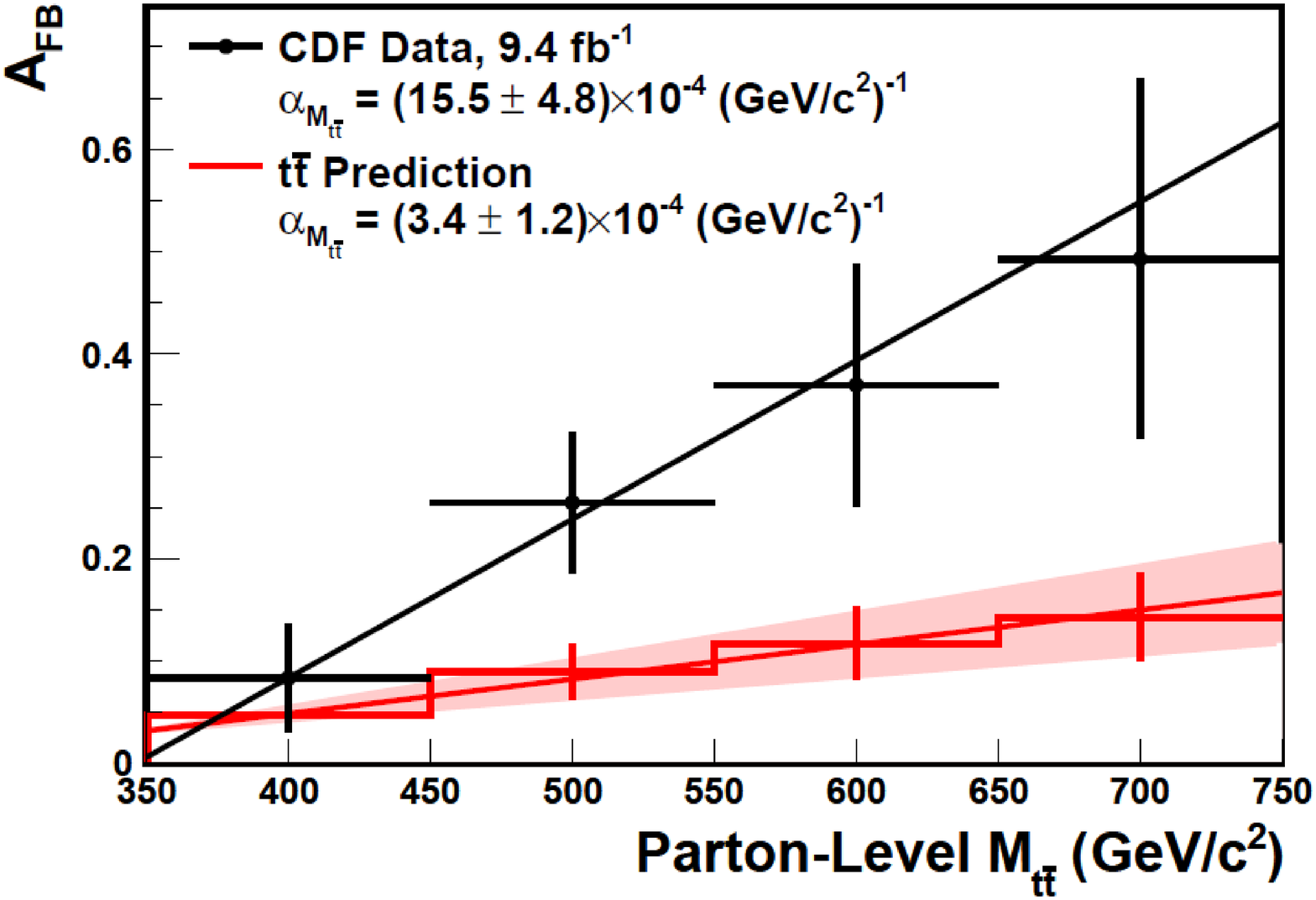}}
\end{minipage}
\hfill
\begin{minipage}{0.48\linewidth}
\centering{\includegraphics[width=0.95\linewidth]{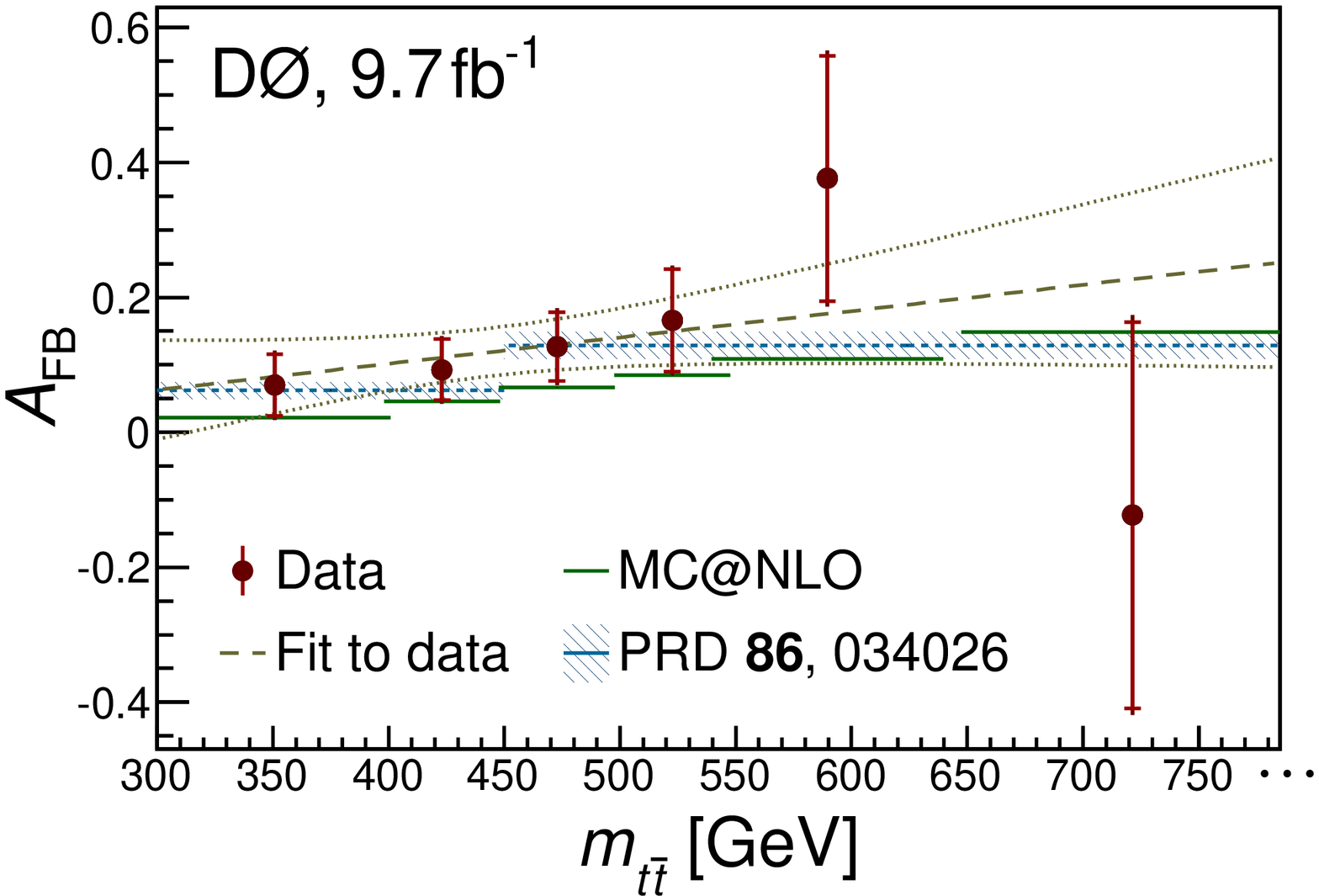}}
\end{minipage}
\label{fig:afb}
\caption{Forward-backward asymmetry as a function of top-quark pair mass at the parton level. The last bin contains overflow events. (left) CDF result, (right) D\O\ result.}
\hfill
\end{figure}

\section{Production asymmetries}\label{sec:asym}
The $t\bar{t}$ pairs are produced through strong force quark-antiquark annihilation or gluon-gluon fusion. In the leading order (LO), the SM does not predict any asymmetry in either of the processes. However, at NLO, the SM predicts asymmetry coming from interference of the amplitudes of Born and box diagrams and interference of the initial and final state gluon radiation in quark-antiquark annihilation processes. Thec$t\bar{t}$ production via gluon-gluon fusion remains symmetric also at the higher orders. Another contribution to asymmetry is due to interference of quark-gluon scattering processes or due to the electroweak interactions, but the expected contribution is small.

\subsection{Forward-backward asymmetry}
After the kinematic reconstruction of final $t\bar{t}$ state, one can define the forward-backward asymmetry using rapidity difference, $\Delta y = y_t - y_{\bar{t}}$:
\begin{equation}
A_{FB} = \frac{N(\Delta y > 0) - N(\Delta y <0)}{N(\Delta y > 0) + N(\Delta y <0)}
\label{eq:afb}
\end{equation}
\noindent
where $y_t$ ($y_{\bar{t}}$) corresponds to rapidity of top (antitop) quark.

Both, CDF and D\O\ measure the asymmetry in the $lepton+jet$ channel using full data sets. After the full kinematic reconstruction of $t\bar{t}$ final state, the correction to the parton level is done using regularized 2D unfolding. The asymmetry measured by the CDF experiment is $A_{FB} = 0.164 \pm 0.039$(stat) $\pm 0.026$(syst)~\cite{CDFafb}, while the value measured by the D\O\ experiment is $A_{FB} = 0.106 \pm 0.027$(stat) $\pm 0.013$(syst)~\cite{D0afb}. 

Both experiments express the inclusive asymmetry as a function of top-quark pair mass, $m_{t\bar{t}}$. In addition to that, CDF presents also dependence of the $A_{FB}$ on the rapidity difference, $|\Delta y|$ and on the transverse momentum of the $t\bar{t}$ system, $p_{T}(t\bar{t})$. The asymmetry at CDF is found to have approximately linear dependence on both $|\Delta y|$ and $m_{t\bar{t}}$, as expected for the NLO charge asymmetry, although with larger slopes then are expected in the NLO prediction. The probabilities to observe the measured values or larger for the detector-level dependencies are 2.8$\sigma$ and 2.4 $\sigma$ for $|\Delta y|$ and $m_{t\bar{t}}$, respectively. The D\O\ measurement of the $A_{FB}$ dependence on $m_{t\bar{t}}$ shows compatibility of the data with the SM predictions as well as with the CDF result (see Fig. 1).

\subsection{Lepton-based asymmetry}

To measure the lepton-based asymmetry, there is no need to reconstruct $t\bar{t}$ final state. The advantage is also a good lepton charge determination and high precision of measurement of the lepton direction. One can define single-lepton asymmetry, $A_{FB}^{l}$, using the lepton charge multiplied by its rapidity ($qy_{l}$); or dilepton asymmetry, $A_{FB}^{\Delta\eta}$, using the difference of pseudo-rapidities of positive and negative leptons ($\Delta\eta = \eta_{l^{+}} - \eta_{l^{-}}$) in dilepton channel. 

Both experiments measure the single-lepton asymmetry in $lepton+jets$~\cite{CDFljafb,D0ljafb} and $dilepton$~\cite{CDFdilafb,D0dilafb} channels and combine the results using the BLUE method. CDF obtains $A_{FB}^{l}=0.090^{+0.028}_{-0.026}$~\cite{CDFdilafb}, while D\O\ measures the value of $A_{FB}^{l}=0.047 \pm 0.023$(stat)$\pm 0.015$(syst)~\cite{D0ljafb}. Comparing the results with the SM prediction of $A_{FB}^{l}=0.038 \pm 0.003$, one can say, that CDF sees $2\sigma$ excess, while D\O\ result is compatible with the expectations. 

In the $dilepton$ channel, CDF and D\O\ measure dilepton asymmetry of $A_{FB}^{\Delta\eta}=0.072 \pm 0.081$~\cite{CDFdilafb} and $A_{FB}^{\Delta\eta}=0.123 \pm 0.054$(stat)$\pm 0.015$(syst)~\cite{D0dilafb}, respectively. Both results are compatible with SM prediction of $0.048 \pm 0.003$. Furthermore, D\O\ presents a ratio of single-lepton and dilepton asymmetries in $dilepton$ channel. There is a discrepancy between the measured value of the ratio of $A_{FB}^{l}/A_{FB}^{\Delta\eta}=0.36\pm0.20$ and the SM prediction of $0.79\pm0.10$.

\section*{Acknowledgments}

It is a pleasure to thank the CDF and D\O\ collaborators for their well-done work, the top-group conveners for their help and the organizers of the Moriond QCD 2014 for a very interesting conference. This work was supported by Ministry of Education, Science, Research and Sport of the Slovak Republic.

\end{document}